\documentclass{eas}
\usepackage{graphicx}
\begin{document}
%================================================================
\title{NEW CCD PHOTOMETRIC OBSERVATION OF THE W-UMA ECLIPSING BINARY SYSTEM 1SWASP J160156.04+202821.6}
%================================================================
\author{A. Essam}\address{National Research Institute of Astronomy and Geophysics, 11421 Helwan,
Cairo - Egypt, e-mail:  essam60@yahoo.com}
\author{Nasser M. A}
%================================================================
\begin{abstract}
\begin{sloppypar}
New \textit{BVRI} light curves of the eclipsing binary system 1SWASP
J160156.04+202821.6 have been obtained with the 1.88 meter telescope
of Kottamia Astronomical Observatory (KAO), Egypt on June, 2013. New
twenty times of minima of the system J1601 and new ephemeris have
been reduced and derived from the present photometry. A preliminary
determination of the geometric and photometric element parameters of
the system J1601 has been derived.
\end{sloppypar}
\end{abstract}
%================================================================
\runningtitle{A. Essam and Nasser M. A.}
\maketitle
%================================================================
%\section{Introduction}
The eclipsing binary system 1SWASP J160156.04+202821.6 (J1601,
$\alpha_{2000}$ = $16^h \;01^m \;56.04^s$, $\delta_{2000}$ = +$20^o
\; 28^{'} \;21.6^{''}$) was discovered as a variable star by Norton
et al. (2011). He reported that the orbital period of the system
J1601 is 0.22653d.
%================================================================
%\section{Observation and data reduction}

Photometric observations of the eclipsing binary system J1601 have
been obtained in {\it B}, {\it V}, {\it R} and I wide pass-band
filters through three nights, $3^{ed}$, $4^{th}$, $5^{th}$ of June,
2013, using EEV CCD 42-40 camera attached to the Newtonian focus of
the 1.88-m telescope of Kottamia Astronomical Observatory (KAO),
Egypt. Differential photometry was performed with respect to GSC
0151100135 and GSC 0151100096 as a comparison and check stars
respectively. All times of the observation were converted to HJD.
The complete light curves in {\it BVRI} bands have been obtained in
the night of $4^{th}$ and $5^{th}$ of June, 2013.
%================================================================
%\section{Epochs of photometric minima}

New twenty times of minima of the system J1601 have been derived
from the present photometry. We used these times of minima to
construct a new ephemeris as follows: (Min I HJD) = 2456448.36291
(�4.14552*$10^{-8}$)    +0.226593 (�1.1436*$10^{-8}$) * E, with E
the integer cycle. We have been used this new ephemeris to phases
our observations.
%=====================================================================
%\section{Preliminary photometric data analysis}

The software Binary Maker 3 (Bradstreet 2005, hereafter BM3) has
been used to construct the syntactic light curves of the {\it BVRI}
pass bands for the system J1601. From the UCAC4 Catalog (see
Zacharias, N. 2012). the (B-V) color index of the system J1601 equal
1.13, which corresponding to effective temperature of the hot
component ($T_h$) approximately equal 4680 $K^o$ (see Flower 1996).
The light curves of the system J1601 reflect a slightly difference
between the two maximums, which mean exhibit a typical O'Connell
effect (O'Connell 1951). The different solutions, with and without
spot on the components were tested with the observations made on
$4^{th}$ June, 2013. The various parameters are then tweaked until
the theoretical light curves perfectly match the observed ones as
shown in Fig. 1. We derived the degree of over contact (fillout) =
22~\%, inclination, $i$ = $81.5{^o}$, temperature, $T_h$ = 4770
$k^{o}$, $T_c$ = 4670 $k^{o}$ , surface potential, $\omega_1 =
\omega_2$ = 3.156404,  mass ratio, $q$ = $M_2 / M_1$ = 0.7. The same
parameters have been employed to display the system configuration at
the phases 0.25, and 0.5 in Fig. 2.
%=====================================================================
\begin{figure}
%\begin{subfigure}{0.25\textwidth}
%  \centering
  \includegraphics[scale=0.18]{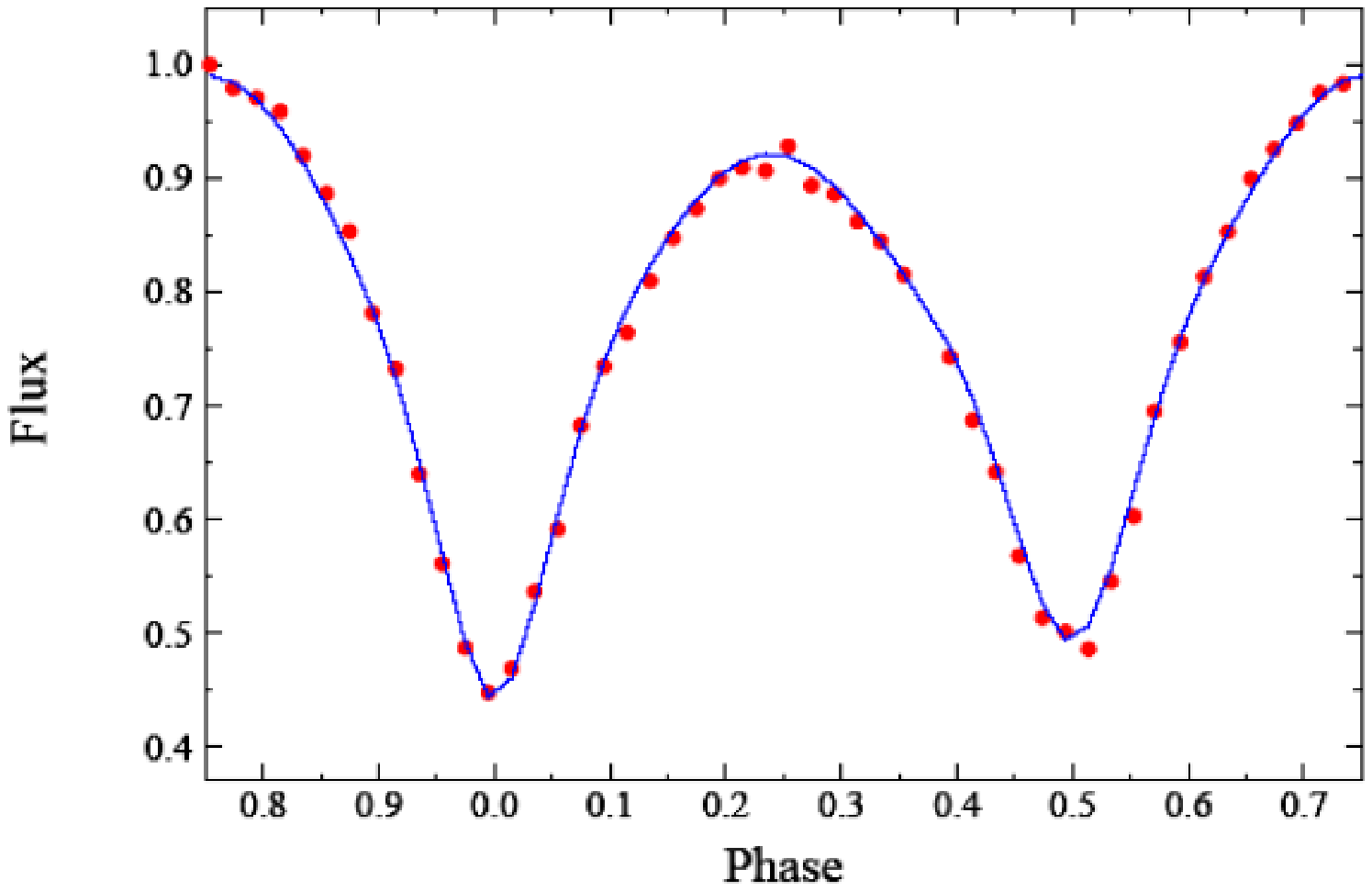}
%  \caption{{\it B-}band}
%  \label{fig:*}
%\end{subfigure}
%===========================================
%\begin{subfigure}{0.25\textwidth}
%  \centering
  \includegraphics[scale=0.18]{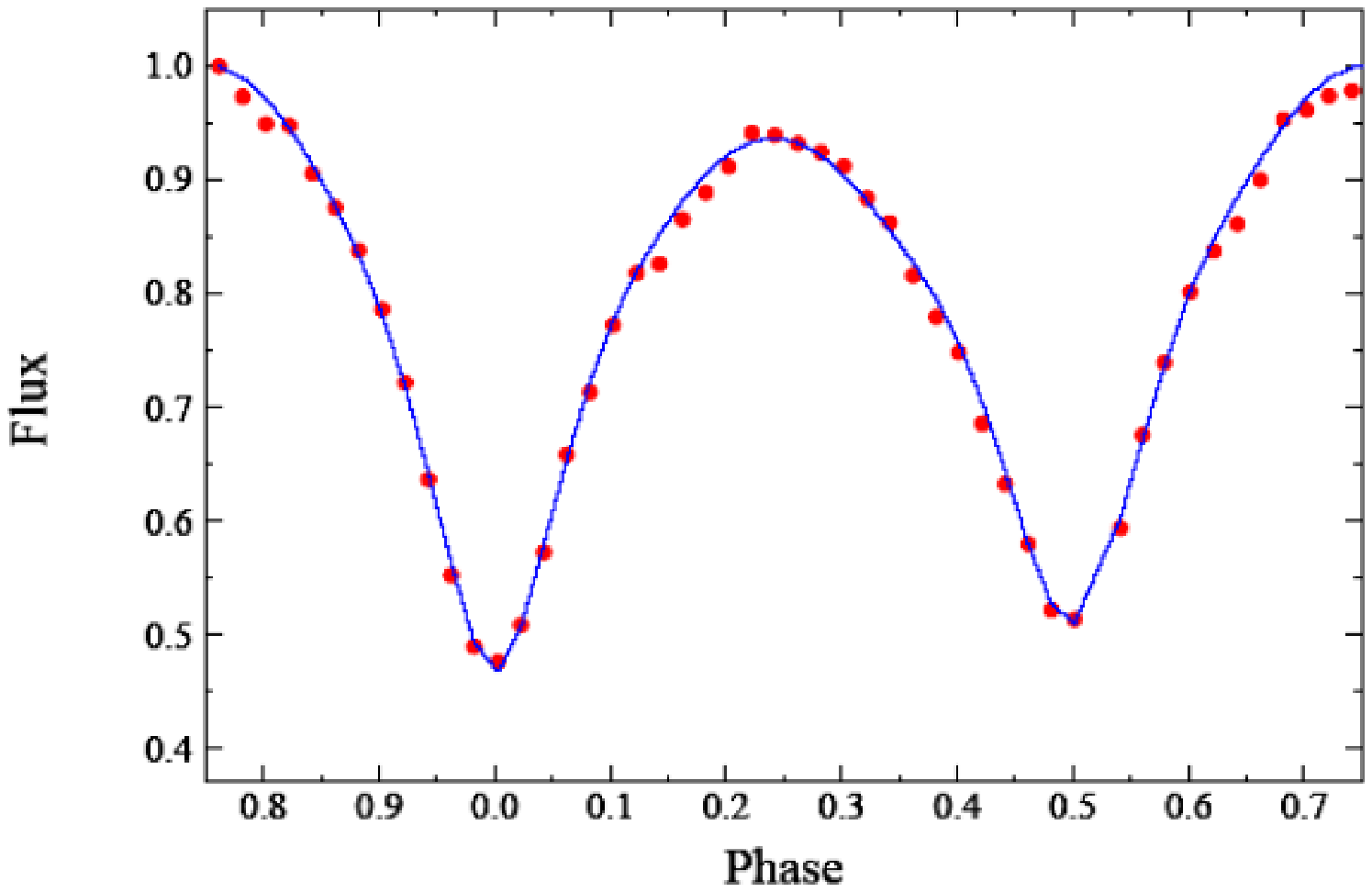}
%  \caption{{\it V-}band}
%  \label{fig:*}
%\end{subfigure}
%============================================
%\begin{subfigure}{0.25\textwidth}
%  \centering
  \includegraphics[scale=0.18]{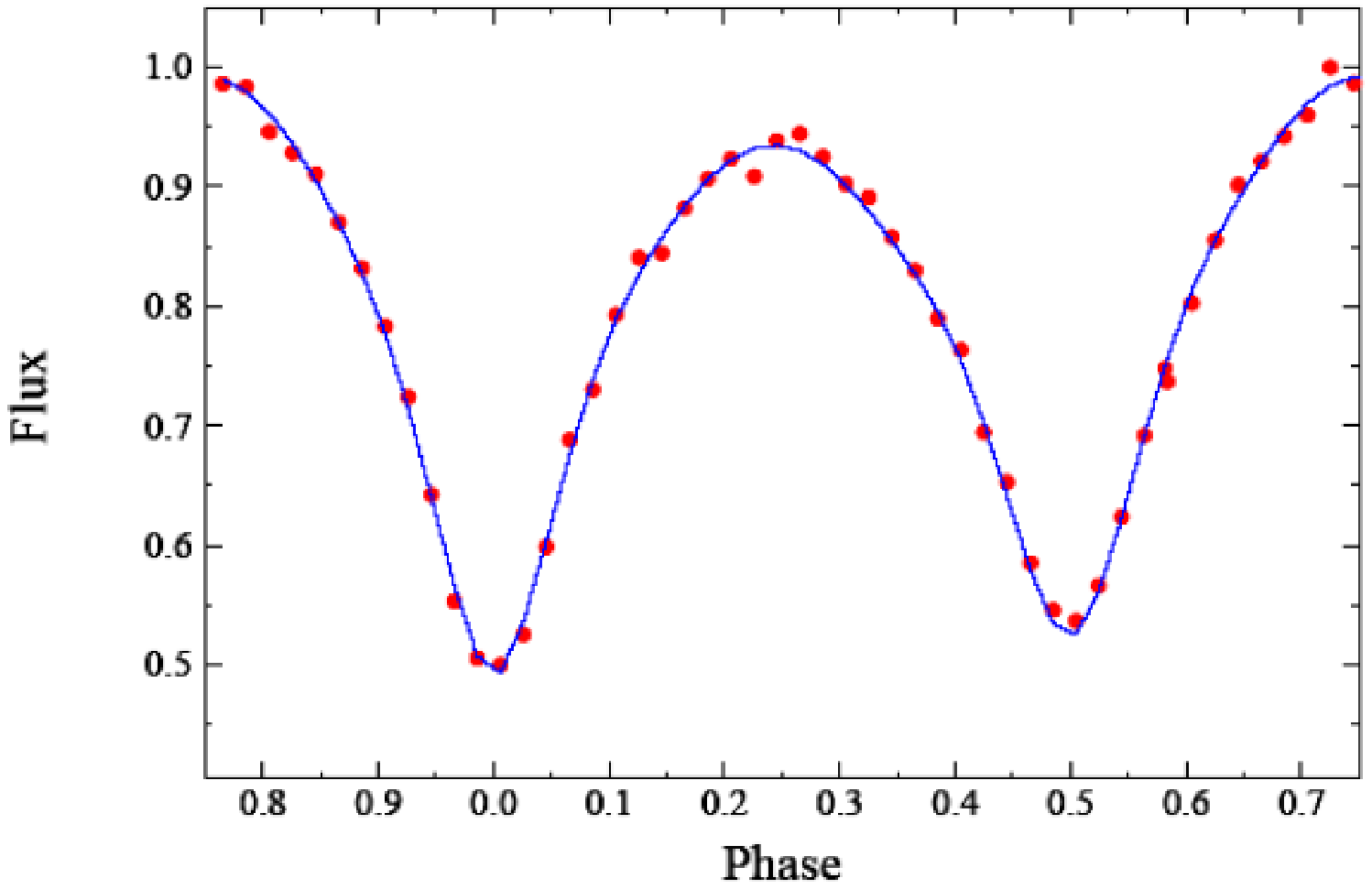}
%  \caption{{\it R-}band}
%  \label{fig:*}
%\end{subfigure}%
%============================================
%\begin{subfigure}{0.25\textwidth}
%  \centering
  \includegraphics[scale=0.18]{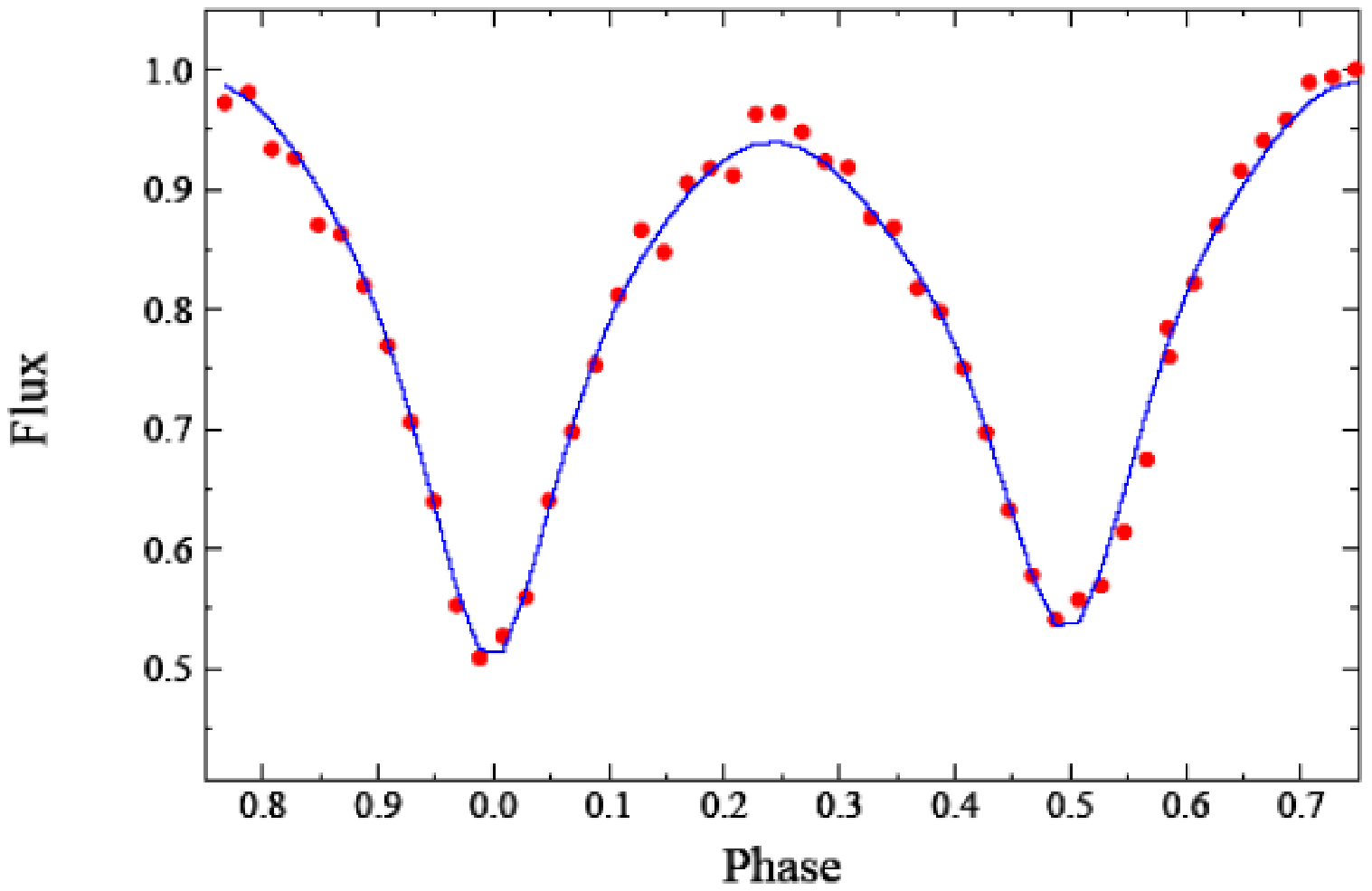}
%  \caption{{\it I-}band}
%  \label{fig:test}
%\end{subfigure}
%============================================
\caption{The {\it BVRI} (from left to right) observed light curves
of the system J1601 (dot points) overlaid with the syntactic light
curve (solid curve).} \label{fig:test}
\end{figure}
%=============================================
\begin{figure}[ht]
%\centering
%=============================================
%\begin{subfigure}{.2\textwidth}
%  \centering
  \includegraphics[scale=0.12,clip=]{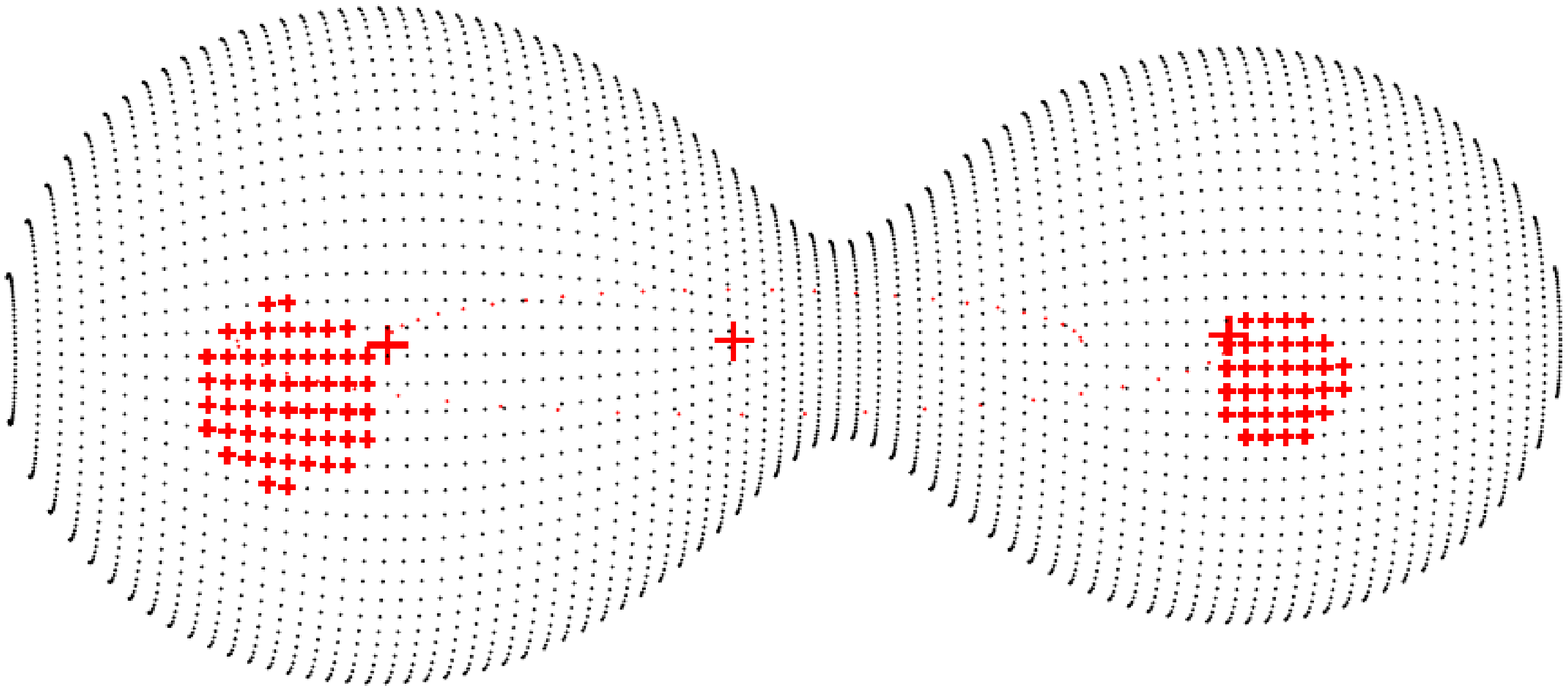}
%  \caption{Phase 0.25}
%  \label{fig:sub1}
%\end{subfigure}%
%=============================================
%\begin{subfigure}{.2\textwidth}
%  \centering
  \includegraphics[scale=0.12,clip=]{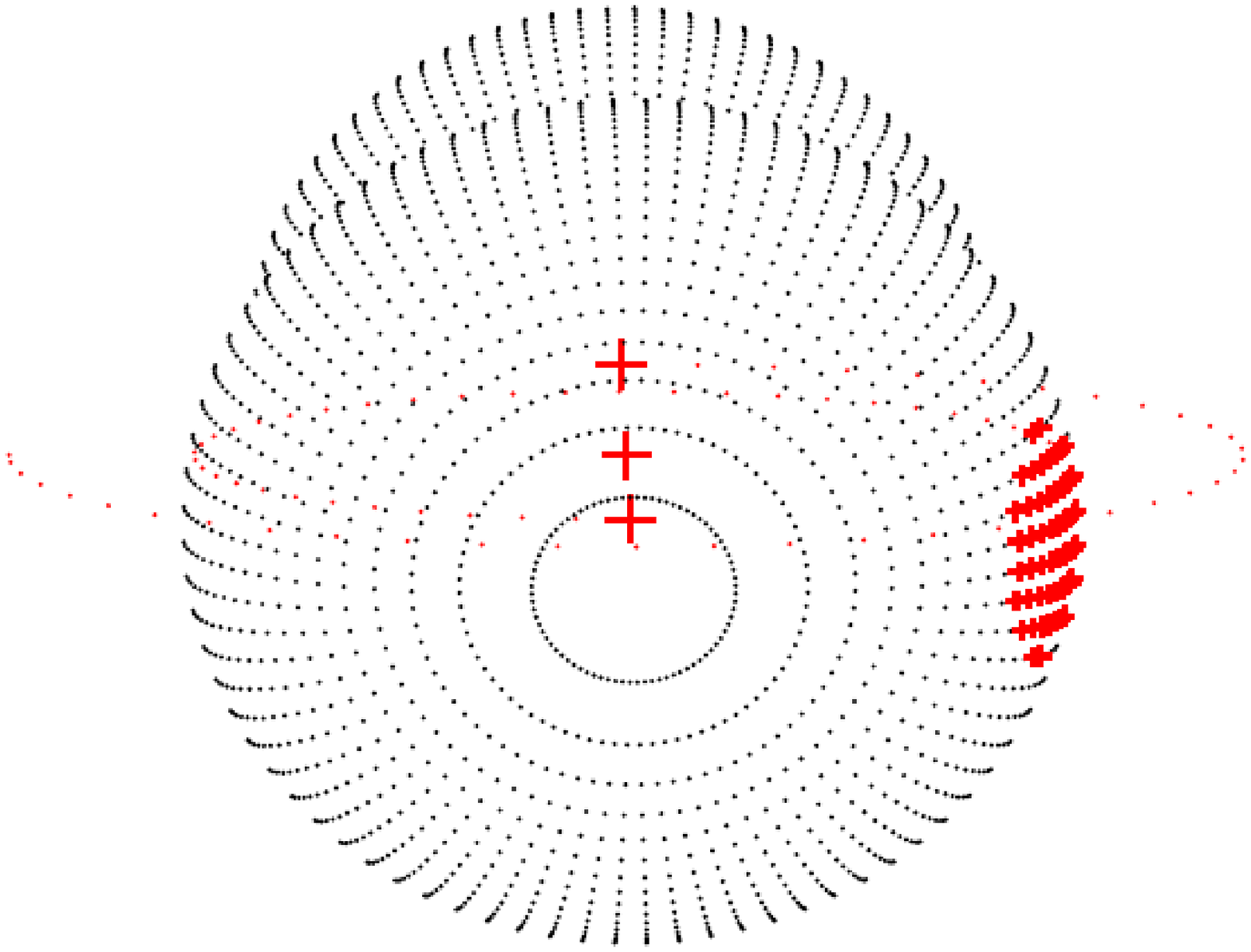}
%  \caption{Phase 0.50}
%  \label{fig:sub2}
%\end{subfigure}
%=============================================
%\begin{subfigure}{.2\textwidth}
%  \centering
  \includegraphics[scale=0.12,clip=]{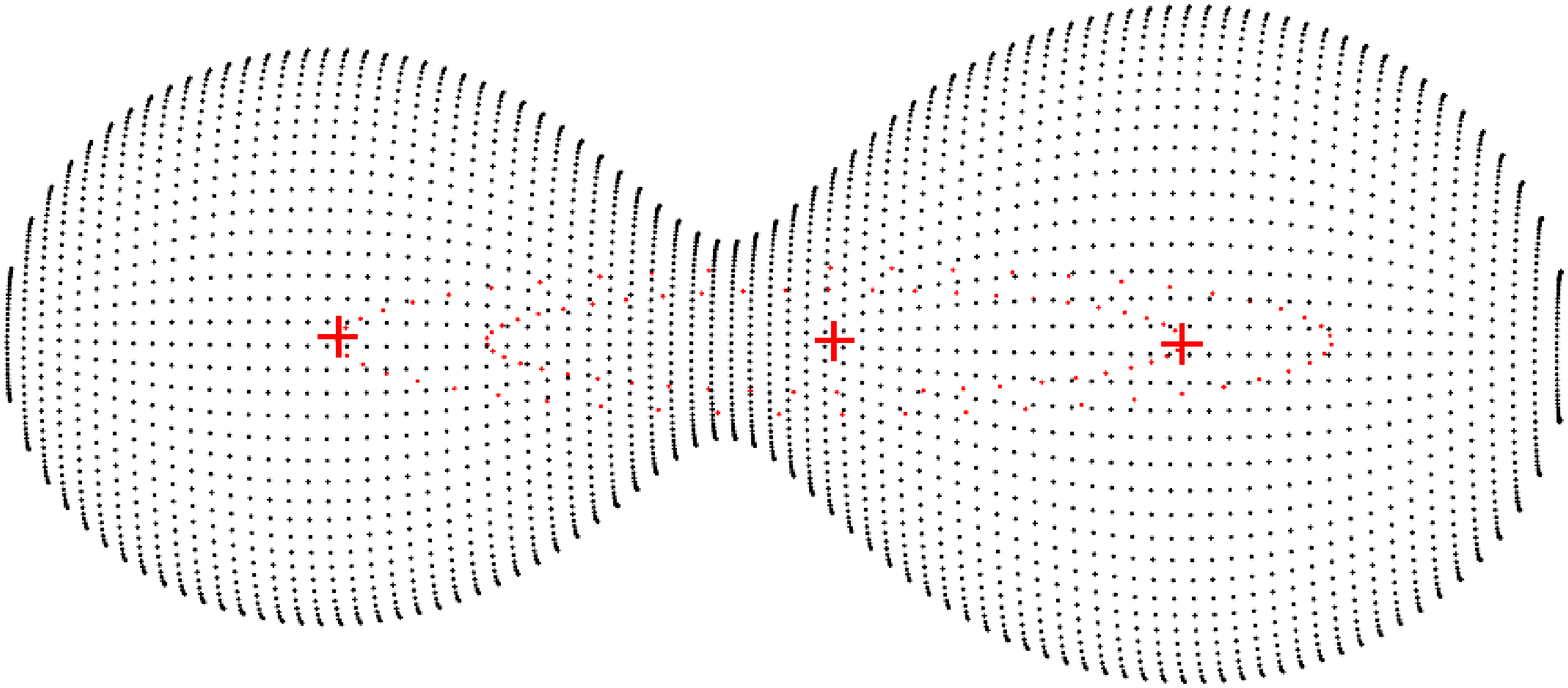}
%  \caption{Phase 0.75}
%  \label{fig:sub2}
%\end{subfigure}
%=============================================
%\begin{subfigure}{.2\textwidth}
%  \centering
  \includegraphics[scale=0.12,clip=]{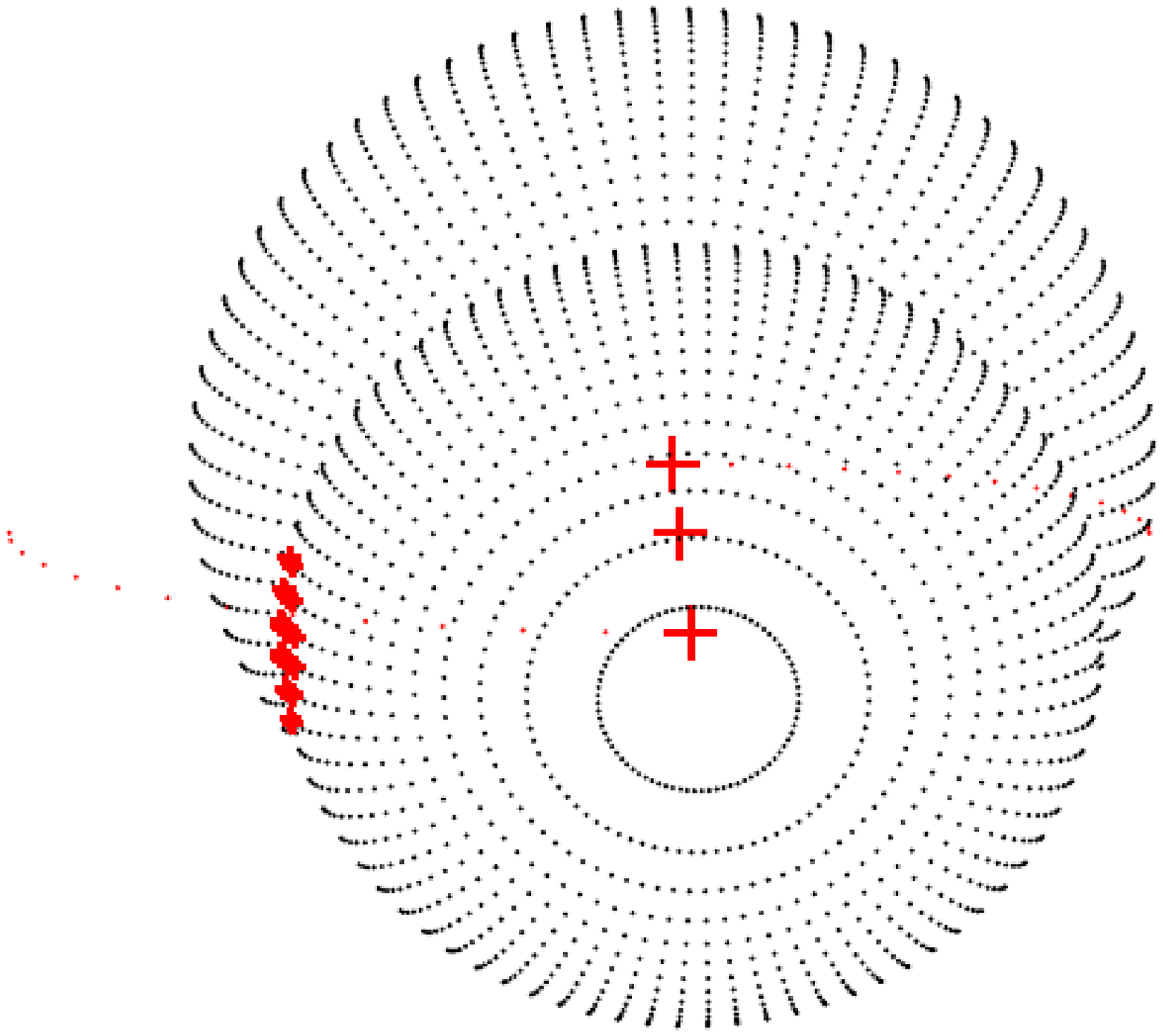}
%  \caption{Phase 1.0}
%  \label{fig:sub2}
%\end{subfigure}
%=============================================
\caption{3D model of the system J1601 at the phases 0.25, 0.5, 0.75,
and 1.0 (from left to right)} \label{fig:test}
\end{figure}
%=============================================================================
%\section{Conclusion}

The {\it B}, {\it R} and {\it I} light curves of the system J1601
have been presented here for the first time, while our {\it V} light
curve is the secondary after that of the Super Wasp. The solution
reveals that the system J1601 is an over contact binary system by
22~\% and follow the spot model. For the contact binary system J1601
the categorization of the system may like to be A sub-type of W-UMa
system, where the primary component star is the hotter and more
massive one.

\section*{Reference}
Bradstreet, D.H.: BINARY MAKER 3.0, Contact Software, 725 Stanbridge Street, Norristown, PA 19401-5505 USA.
http://www.binarymaker.com/ (2005).\\
Flower, P. J.: The Astronomical Journal, 469: 355-365, 1996 \\
Norton, A.J., Payne, S.G., Evans, T., et al.: Astron. Astrophys. 528, 90 (2011)  \\
O'Connell D. J. K., 1951, Pub. Riverview College Obs., 2, 85 \\
Zacharias, N., Finch, C. T., Girard, T.~M., et al. 2012, VizieR
Catalog, 1322, 0
\end{document}